\documentclass[aps,prl,floats,twocolumn,showpacs,superscriptaddress]{revtex4}

\usepackage{graphicx}
\usepackage{times}
\usepackage{graphics,dcolumn,bm,fleqn,epic,eepic,float}
\usepackage{amssymb,amsmath,multirow,rotate,color}
\usepackage{subfigure}
\begin{document}

\title{Traffic optimization in transport networks based on local routing}

\author{Salvatore Scellato}
\affiliation{Laboratorio sui Sistemi Complessi, Scuola Superiore di Catania,
Via San Nullo 5/i, 95123 Catania, Italy}

\author{Luigi Fortuna}
\affiliation{Dipartimento di Ingegneria Elettrica, Elettronica e dei Sistemi,
Universit\`a degli Studi di Catania, viale A. Doria 6, 95125 Catania, Italy}
\affiliation{Laboratorio sui Sistemi Complessi, Scuola Superiore di Catania,
Via San Nullo 5/i, 95123 Catania, Italy}

\author{Mattia Frasca}
\affiliation{Dipartimento di Ingegneria Elettrica, Elettronica e dei Sistemi,
Universit\`a degli Studi di Catania, viale A. Doria 6, 95125 Catania, Italy}
\affiliation{Laboratorio sui Sistemi Complessi, Scuola Superiore di Catania,
Via San Nullo 5/i, 95123 Catania, Italy}

\author{Jes\'us G\'omez-Garde\~nes}
\affiliation{Laboratorio sui Sistemi Complessi, Scuola Superiore di Catania,
Via San Nullo 5/i, 95123 Catania, Italy}
\affiliation{Departamento de Matem\'atica Aplicada, ESCET, Universidad
Rey Juan Carlos, E-28933 M\'ostoles (Madrid), Spain}

\author{Vito Latora}
\affiliation{Dipartimento di Fisica e
    Astronomia, Universit\`a di Catania, and INFN, Via S. Sofia 64, 95123
    Catania, Italy}
\affiliation{Laboratorio sui Sistemi Complessi, Scuola Superiore di Catania,
Via San Nullo 5/i, 95123 Catania, Italy}

\date{\today}

\begin{abstract}
Congestion in transport networks is a topic of theoretical
interest and practical importance. In this paper we study the flow
of vehicles in urban street networks. In particular, we use a
cellular automata model to simulate the motion of vehicles along
streets, coupled with a congestion-aware routing at street
crossings. Such routing makes use of the knowledge of agents about
traffic in nearby roads and allows the vehicles to dynamically
update the routes towards their destinations. By implementing the
model in real urban street patterns of various cities, we show
that it is possible to achieve a global traffic optimization based
on local agent decisions.
\end{abstract}

\pacs{89.20.-a, 89.75.-k, 89.40.Bb}

\maketitle

{\em Introduction.-} Traffic optimization has always been a crucial
issue in the context of communication and transportation systems
\cite{wardrop,helbing01,chowdhury00,vesp,jentsch}. A transport network
is a network of roads, streets, pipes, power lines, or nearly any
structure which permits either vehicular movement or the flow of some
commodity. In most of the developed countries, transportation
infrastructures originally designed to carry a defined amount of
traffic are often congested by an overwhelming request of resources:
this is the case of railroads, airplane connections and, of course,
urban streets. A naive solution to the problem consists in expanding
the infrastructure to match the increasing demand. However, this is
not always possible due to limitations in available space, negative
outcomes or shortness of resources. A better approach is to carefully
tune the behavior of the existing infrastructures to efficiently
exploit their actual structure and accommodate the new traffic
demands.

The study of congestion in complex networks has been mainly focused on
information systems, such as grid-computing networks or the Internet
\cite{vesp,physrep}.  In such context, diverse solutions have been
proposed in order to increase the network load avoiding the onset of
congestion \cite{arenas}.  In particular, congestion-aware routing, in
which the nodes of the network (the routers) redirect dynamically the
information packets across the less congested paths, has proved to
improve notably the capacity of the network
\cite{echenique1,echenique2}.  It thus seems possible
to make use of similar kinds of routing strategies in transport networks.
On the other hand, transport networks present important features
different from information systems. First, in a transport network the
links ({\em i.e.} the roads) carry the flow of vehicles, whereas the
nodes are just intersections between links.  Therefore, one cannot
neglect the dynamics that occurs along the links: the quality of
vehicle movement along the roads characterizes the functioning of
transport systems. Another important feature is that, since transport
networks are embedded in the real space, congestion is not located at
particular nodes of the system (such as the hubs in information
systems) \cite{clm04a} but it geographically spreads across the
network from bottlenecks and it may eventually affect a large portion
of the system.  Therefore, to study congestion-aware routing of
vehicular traffic in transport networks it is necessary to incorporate
the above two ingredients.

In this paper, we focus on the realistic scenario in which only a
{\em local knowledge} of congestion is available and used by the
agents to modify their routes. Our model is implemented in terms
of vehicular traffic and we assume that drivers know the shortest
paths to their destinations and, simultaneously, they are aware
about the congestion of nearby roads. Both informations are easily
accesible nowadays from navigating systems, visual inspection and
short-range wireless communication with other vehicles
\cite{leontiadis}. We show how, by conveniently controlling agent
decisions, it is possible to minimize the overall congestion of
the system and achieve performances as good as those obtained with
complete (global) traffic information.

\smallskip

{\em The model.-} The three ingredients of the model are:
{\em i)} the substrate graph,
{\em ii)} the vehicular dynamics along the streets,
{\em iii)} the routing at street crossings.

The dynamics of vehicles takes place on top of urban graphs. We
consider the street pattern of a city as a weighted graph with $N$
nodes and $K$ edges. Each edge of the network represents a street,
along which vehicles move, whereas nodes account of intersections
between streets. The weight of each edge is proportional to the length
of the road \cite{urban}. Here we assume for simplicity that each edge
allows movement of vehicles in both directions. We have considered
eight networks representing $1$-square mile samples of urban street
patterns of real cities \cite{urban} (see Table \ref{table} for
details). The above city set ranges from {\em self-organized} cities,
such as Bologna and London, grown through a continuous process out of
the control of any central agency, to {\em grid-like} cities such as
Los Angeles and Barcelona, realized over a short period of time as the
result of urban plans, and usually exhibiting grid structures.

The vehicular dynamics along the links of the urban graph is simulated
by means of cellular automata \cite{wolfram83,esser97,helbing98}. In
particular we use the {\em Nagel-Schreckenberg} (NaSch) model
\cite{nasch}. To this end, every link (street) of the city graph is
divided into a sequence of cells of equal length ($5$ meters) so that
no more than one vehicle can occupy a cell at every time step. Each
vehicle is assigned a velocity $v$ cells per time step \cite{timestep}
in the range $v\in [0,v_{max}]$. We set $v_{max}=3$, corresponding to
the typical maximum velocity of $50$ Km/h inside urban
areas. Following the NaSch model, vehicles accelerate (decelerate)
when the next cells are empty (occupied). Additionally, the
intersections between streets (the nodes of the graph) also allow only
one vehicle at a given time, so that several vehicles coming from
different adjacent streets may compete for the same intersection
\cite{biham92}. For this reason, vehicles experience a slow down while
approaching a road intersection, as if the end of the lane presented a
hindrance, so that they arrive at the last cell of the edge with
$v=0$. At this point, a vehicle waits to enter into the node
\cite{lattice} where it gets stuck until the first cell of the edge in
the proper outgoing direction is free. This waiting locks the incoming
flows from the edges. Therefore, bottlenecks are created from the
nodes and spread along the edges (roads) of the graph.

How vehicles decide their outgoing direction when leaving a node? Here we
implement a {\em congestion-aware} routing as a minimization problem that
takes into account the length of the path and also the traffic along the
outgoing edges.  When a vehicle is at a node $i$ it needs to choose a new node
$n$ in its neighborhood $\Gamma_i$ as the next hop on its path towards the
destination $t$. For each of the neighbouring nodes $n$, a \textit{penalty
function} $P_n$ is defined as:
\begin{equation}
 P_n = (d_{in} + d_{nt}) (1+c_{in})^\alpha\;,  ~~~ n  \in  \Gamma_i
\label{eq:penalty}
\end{equation}
where $d_{in}$ is the distance between nodes $i$ and $n$, and $c_{in}$
($c_{in} \in [0,1]$) represents the congestion of the link $i \rightarrow n$,
measured as the fraction of occupied cells in the link. The exponent $\alpha
\geq 0$ accounts for the weight given to the local congestion in the drivers
decision. The vehicle chooses the node $n$ with the minimum penalty $P_n$.  If
$c_{in}=0$ the penalty function $P_n$ is nothing else than the lenght of the
shortest path to $t$, passing by node $n$. When $c_{in} \neq 0$ the entire
shortest path length is corrected by the factor $(1+c_{in})^\alpha$. In this
way, we assume that the vehicle projects the congestion $c_{in}$ of the link
$i \rightarrow n$ on the entire path $i \rightarrow n \rightarrow t$.  Note
that, when $\alpha=0$, local congestion plays no role in the routing and
vehicles follow the shortest paths to their respective destinations.

\begin{table}[htbp]
\centering
\begin{tabular}{|c|c|c|c|c|c|c|}
\hline
City & $N$ &  $K$ & $W$  & $\left\langle l \right\rangle$  & $\alpha^*(L=0.06)$ & $\alpha^*(L=0.1)$\\
\hline
\hline
Barcelona & 210 & 323 & 36179 & 112.01 & 0.63 & 2.04\\
\hline
Bologna & 541 & 773 & 51219 & 66.26 & 0.87  & 1.54 \\
\hline
Brasilia & 179 & 230 & 30910 & 134.39 & 1.52  & 2.09 \\
\hline
Los Angeles & 240 & 340 & 38716 & 113.87 & 1.82 & 2.43 \\
\hline
London & 488 & 730 & 52800 & 72.33 & 1.02 & 1.26 \\
\hline
New Delhi & 252 & 334 & 32281 & 96.56 & 1.49 & 2.21\\
\hline
New York & 248 & 419 & 36172 & 86.33 & 0.74 & 1.77\\
\hline
Washington & 192 & 303 & 36342 & 119.94 & 1.33  & 2.14\\
\hline
\end{tabular}
\caption{City samples considered: $N$ and $K$ are number of nodes and links in
  the network, $W$ and $\left\langle l \right\rangle$ are respectively the
  total length of edges and the average edge length (both in meters)
  \cite{urban}. We report the value $\alpha^*$ that maximizes the average
  number of completed routes per vehicle.}
\label{table}
\end{table}

We initially place a number of vehicles proportional to the number of cells in
the network, so that the network {\em load}, $L = \frac{V}{C}$ ({\em i.e.} the
ratio between the number of vehicles $V$ in the network and the total number
of cells $C$), is the same for all the cities considered. Initially, the
vehicles are assigned a random source (their initial location) and a random
destination node.  At each time step, vehicles move (if possible) in the
system according to the NaSch rules and the congestion-aware routing,
Eq. (\ref{eq:penalty}). Finally, when a vehicle reaches its destination it is
randomly assigned to a new destination node, so that $L$ is constant in time.
After an initial transient dynamics, the system reaches a steady state in
which data is collected.

\begin{figure*}[htbp]
      \subfigure[]
      {\fbox{\includegraphics[width=0.65\columnwidth]{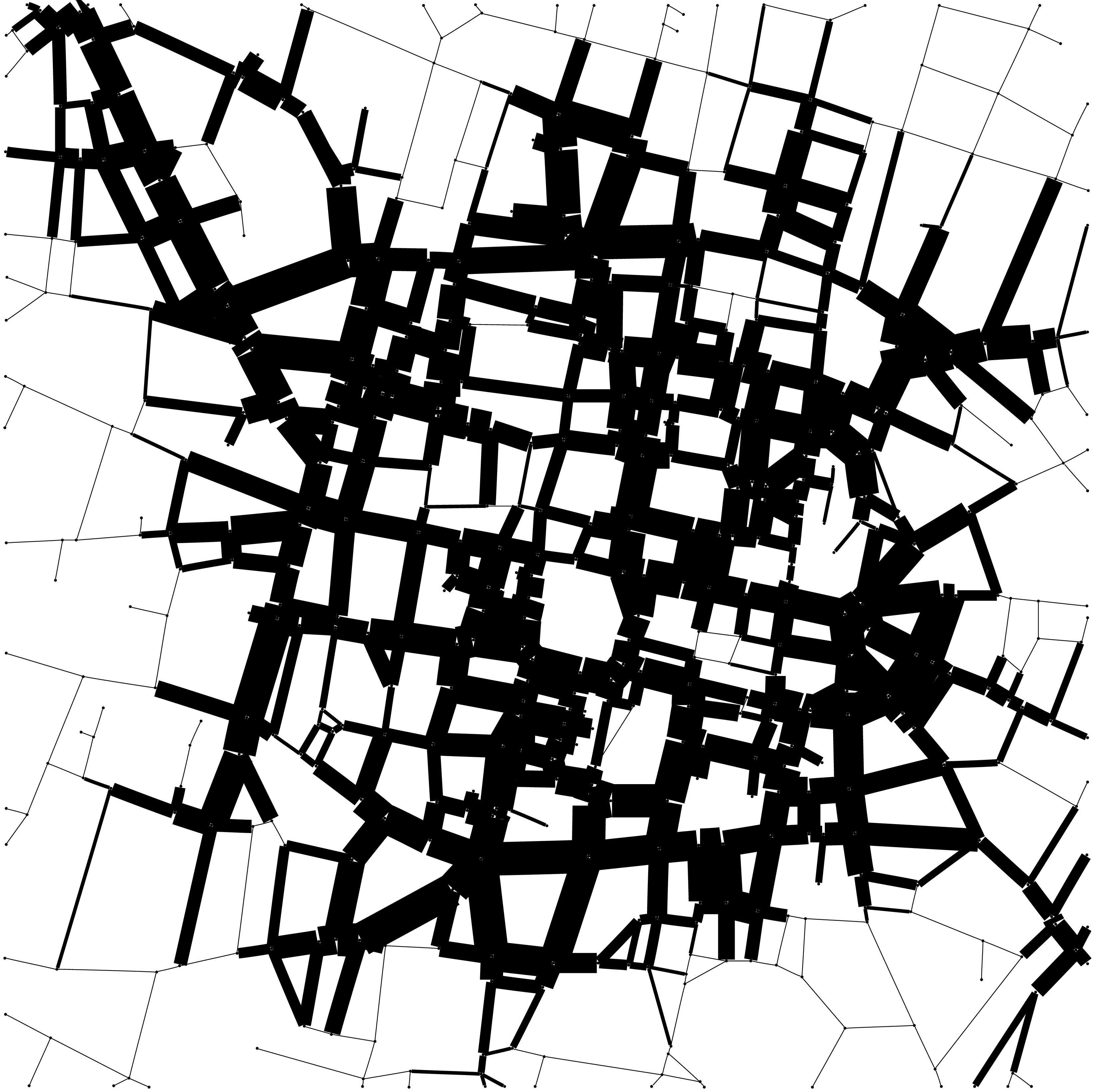}}}
      \subfigure[]
      {\fbox{\includegraphics[width=0.65\columnwidth]{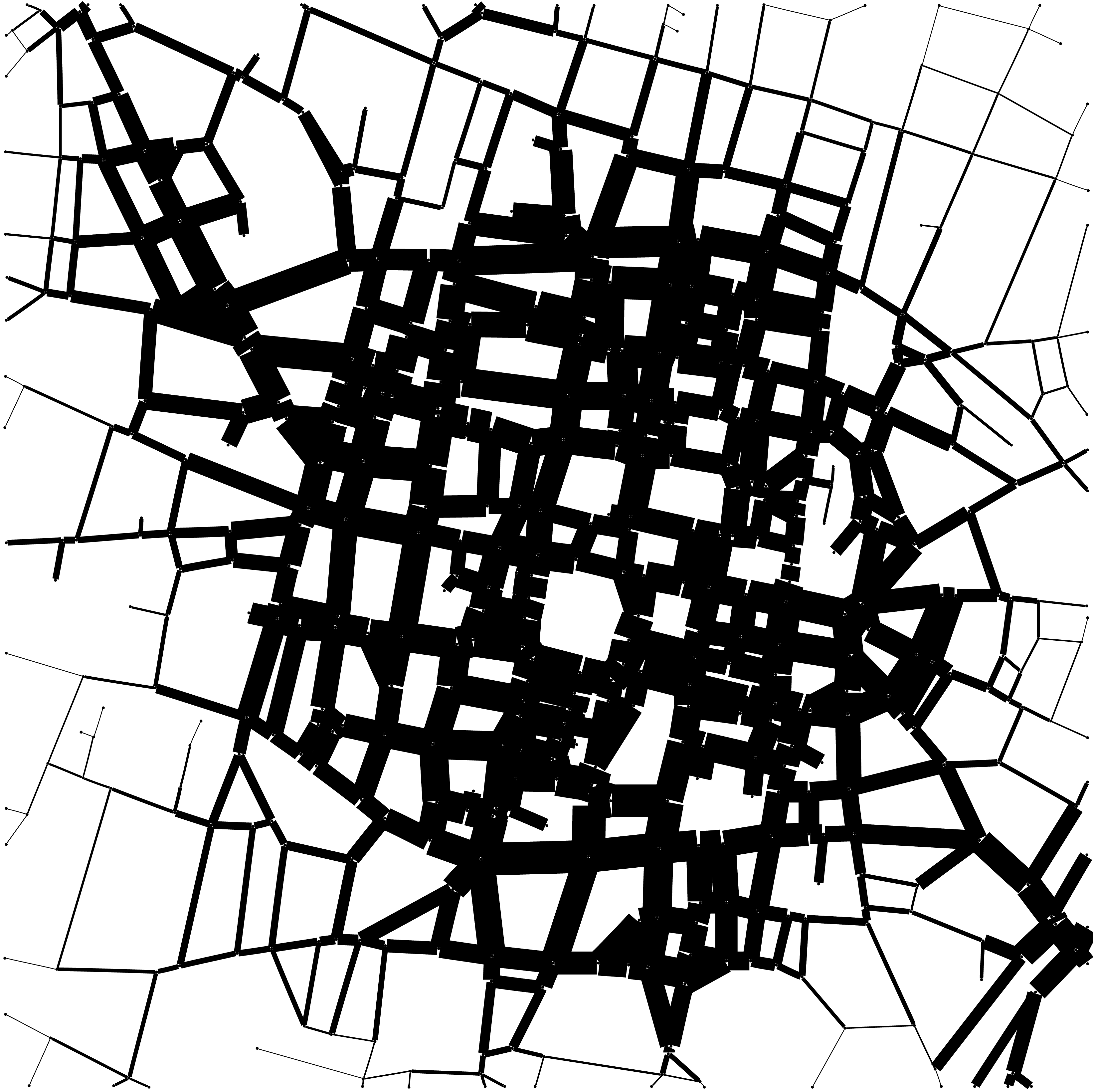}}}
      \subfigure[]
      {\fbox{\includegraphics[width=0.65\columnwidth]{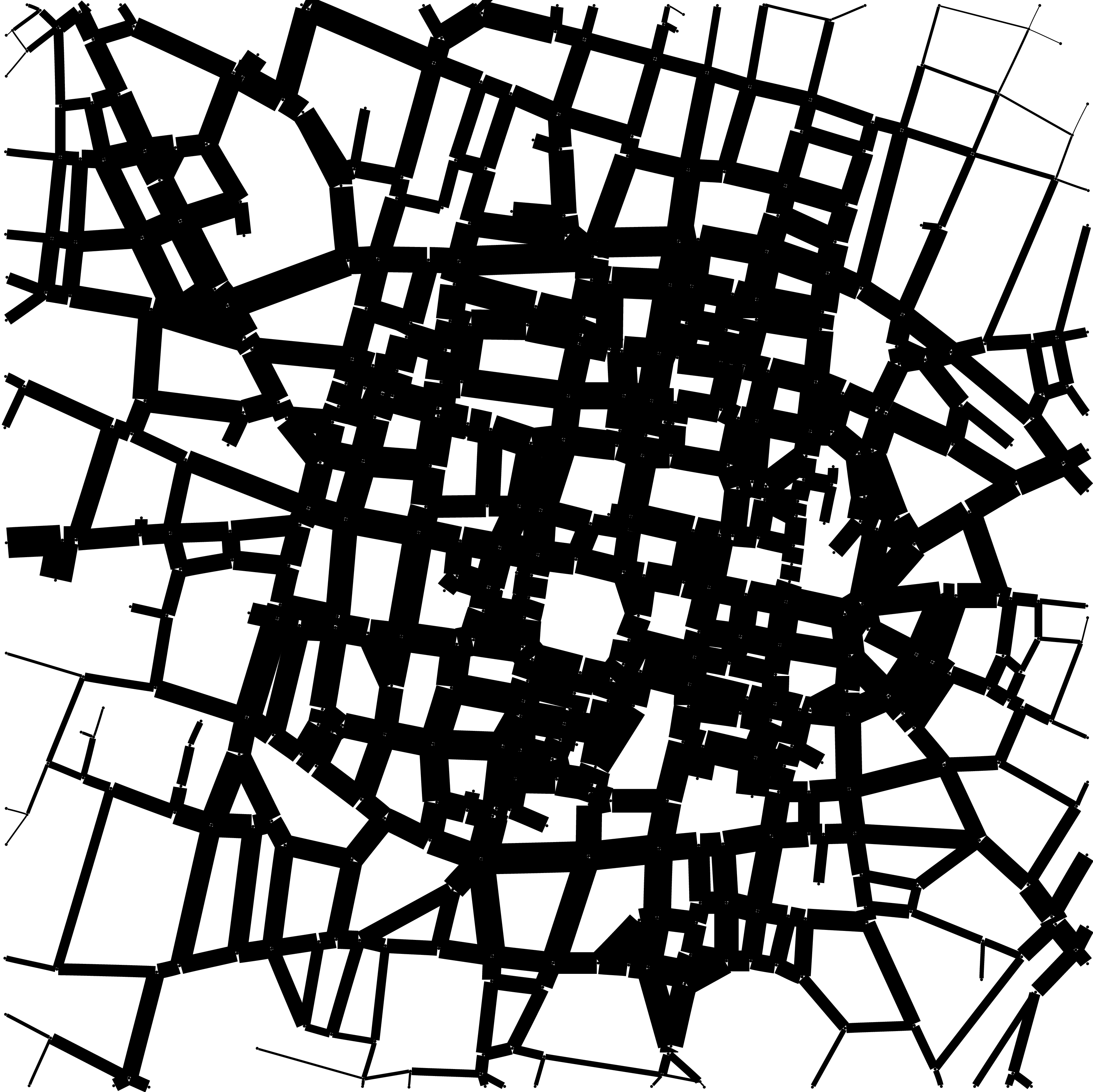}}}
    \caption{Urban street network of Bologna (one-square-mile sample).
Links are drawn with a thickness proportional to their congestion $c$:
from (a) to (c) a more congestion-aware
strategy rules the same amount of traffic, which progressively flows
in a larger number of streets. The traffic load is fixed at L= 0.2,
while the values of $\alpha$ considered are respectively equal to 0,
1 and 2.}
    \label{fig:trafficMapsLinks}
\end{figure*}

\smallskip

\begin{figure*}[htbp]
     \subfigure[]
      {\includegraphics[width=0.6\columnwidth]{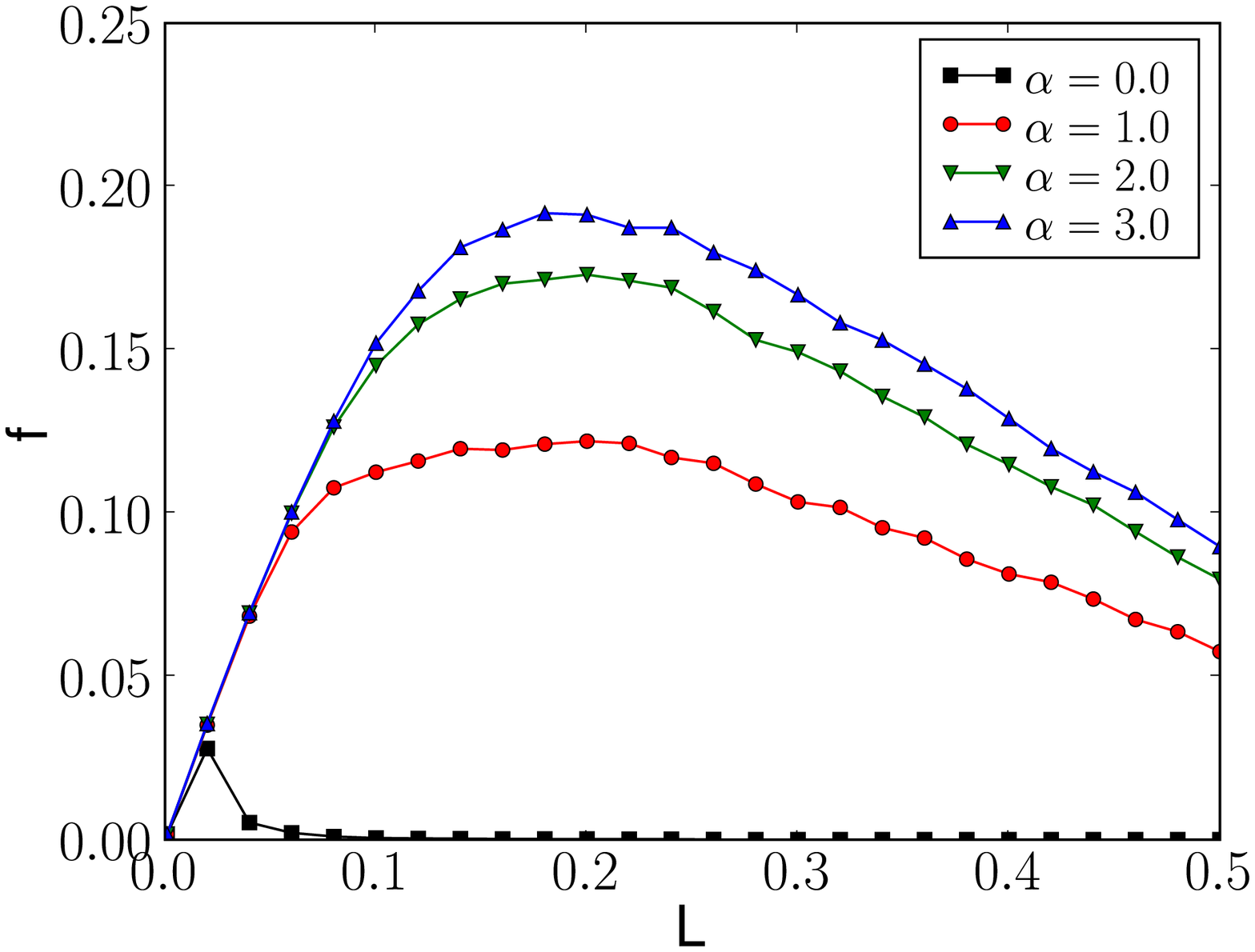}}\hfil
      \subfigure[]
      {\includegraphics[width=0.6\columnwidth]{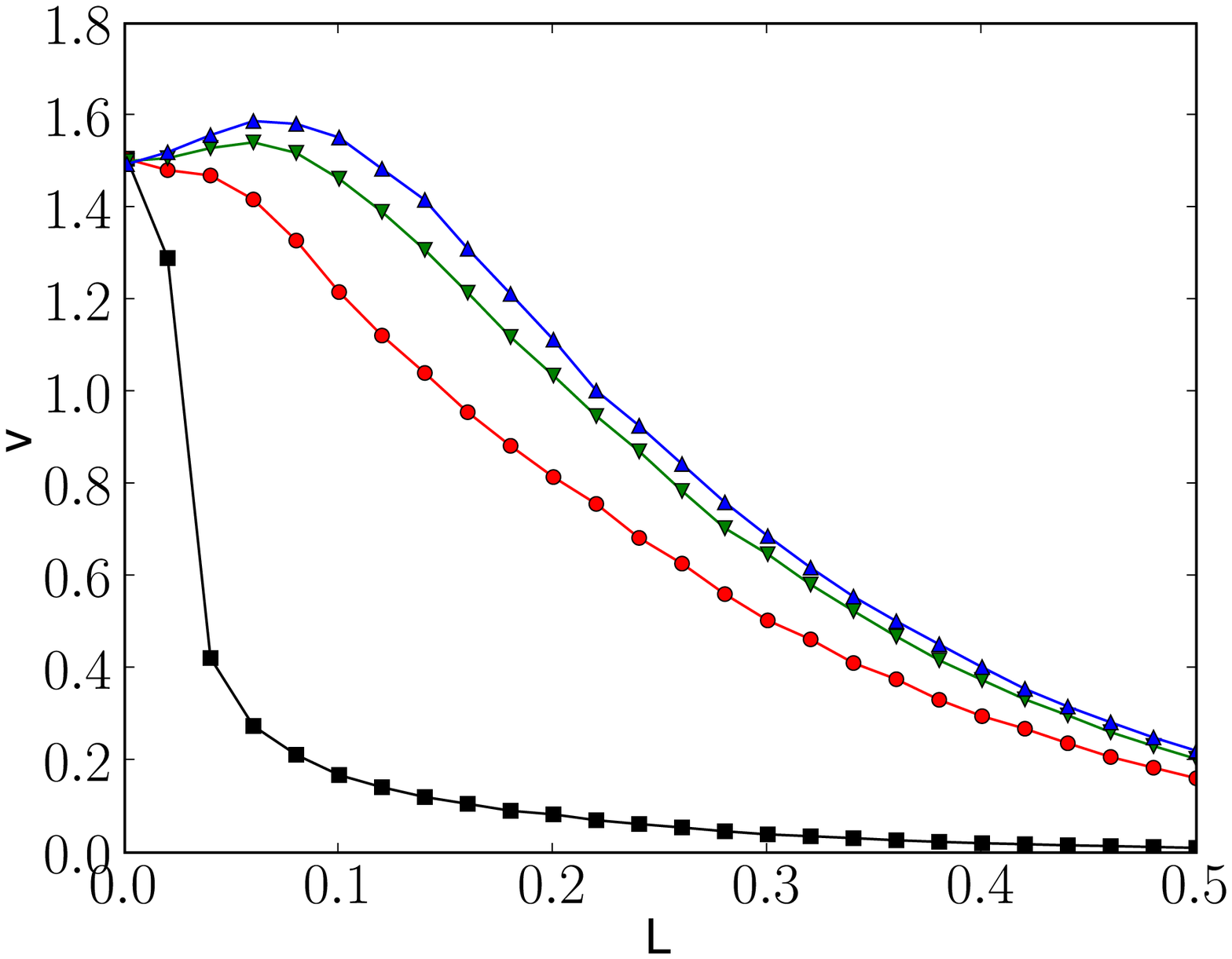}}\\
      \subfigure[]
      {\includegraphics[width=0.6\columnwidth]{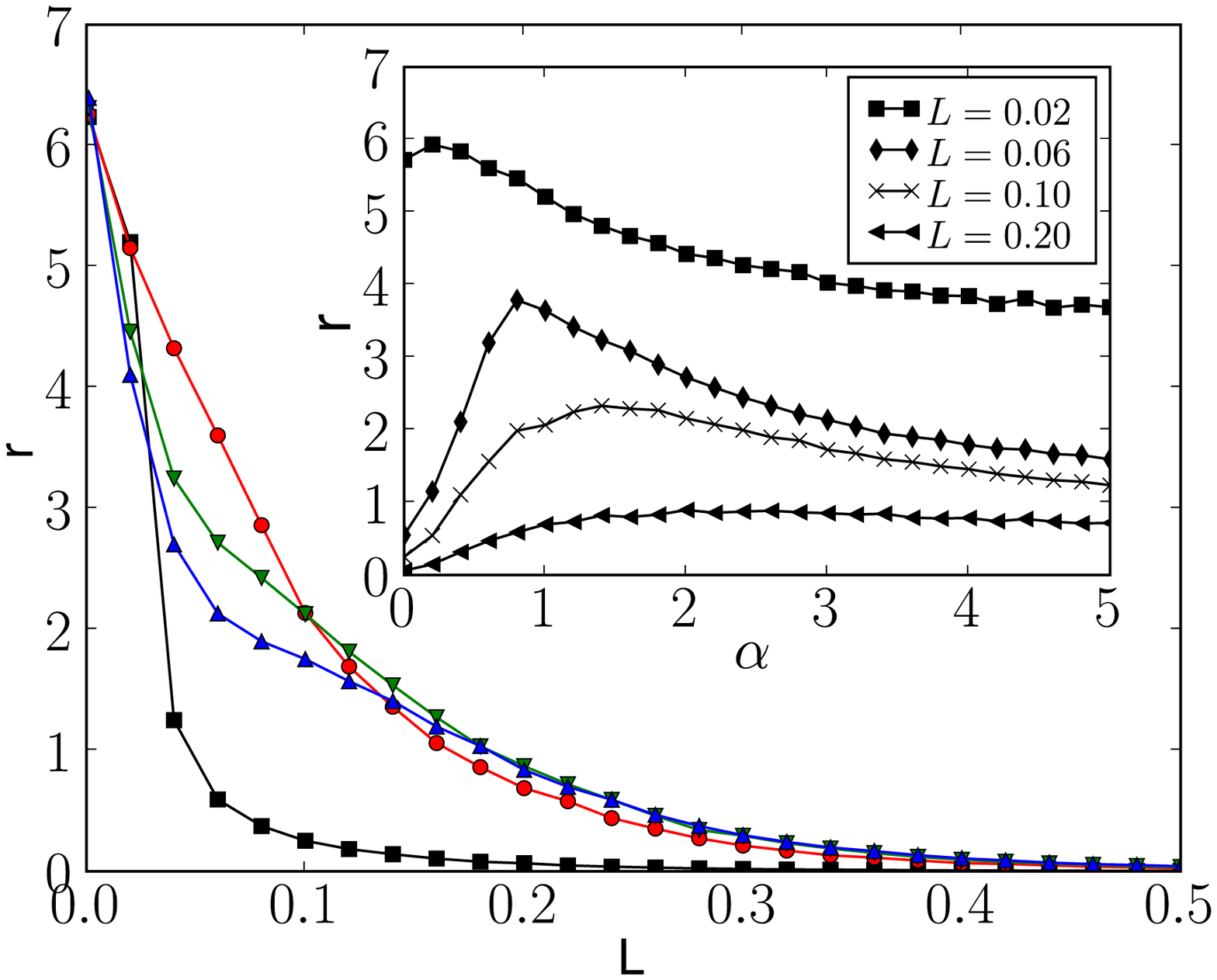}}\hfil
      \subfigure[]
      {\includegraphics[width=0.6\columnwidth]{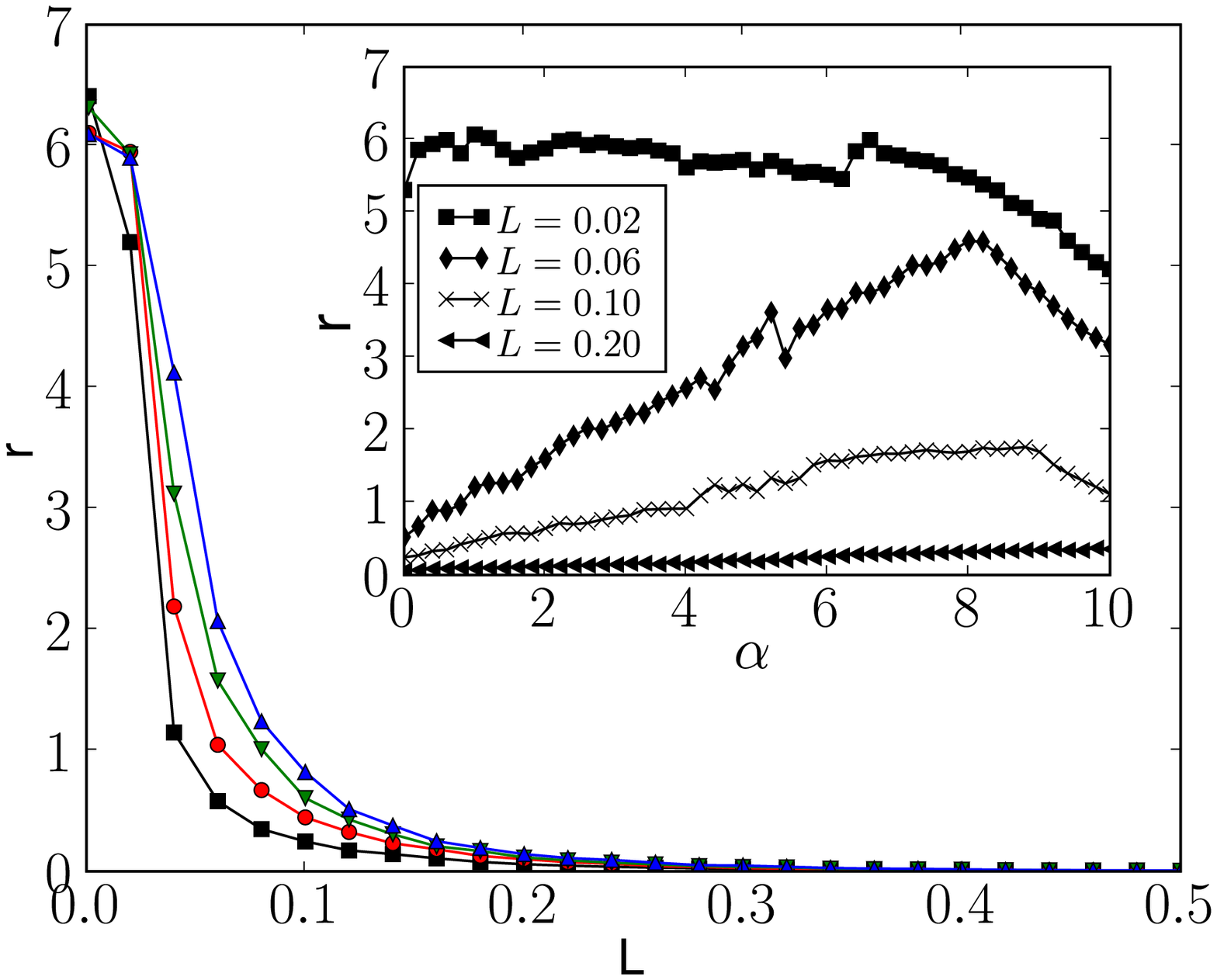}}
\caption{Average street flow (a), average vehicle speed (b), and average
number of completed routes per vehicle per hour (c) as a function of the
network load $L$. The average number of completed routes per vehicle for the
global-aware strategy is also reported for comparison in panel (d).  In the
insets we report $r$ as function $\alpha$.  The network considered is that of
the city of Bologna.  }
\label{fig:fundamental}
\end{figure*}

{\em Results.-} Now we report the dynamical behavior of the model in
the different cities considered as a function of the routing,
$\alpha$, and the network load, $L$. First, in
Fig. \ref{fig:trafficMapsLinks} we show the congestion pattern across
the streets of Bologna for a load $L=0.2$, and for three different
values $\alpha=0$, $1$, $2$. It is clear that the larger the value of
$\alpha$, the more homogeneously distributed is the traffic. In order
to study and quantify the effects of congestion in the vehicular
motion we analyze the so-called {\em fundamental diagram}
\cite{nasch}. The fundamental diagram represents the network mean flux
$f$ (the average value of fluxes on streets) as a function of the
traffic load $L$. This is shown in Fig.~\ref{fig:fundamental}(a) for
different values of $\alpha$. Each of the curves $f(L)$ exhibits the
typical $\lambda$-inverse shape with the two traffic phases: {\em
  free-flow} (in which the flow increases as a function of the load)
at low $L$, and congested-flow (the flow decreases as a function
of the load) at higher values of $L$.  Interestingly, both the
value of maximum flow and the transition point from the free-flow
regime to the congested-flow regime, increase with $\alpha$.  In
particular, when the vehicles follow shortest paths ($\alpha=0$),
they tend to concentrate in high betweenness nodes and links (we
have checked that the strongest correlation between link flow and
betweenness is indeed obtained for $\alpha=0$).  In this case,
even a small density of vehicles can result in a heavy load at
high betweenness streets. This gives rise to the formation of
clusters of jammed vehicles that cannot easily move and, although
large regions of the city are quite uncongested as shown in
Fig.~\ref{fig:trafficMapsLinks}(a), the overall flux of the
network is largely reduced.

The above result is confirmed by the sharp decrease of the average vehicle
speed $v$ as a function of $L$ for $\alpha=0$ shown in
Fig.~\ref{fig:fundamental}(b).  On the other hand, as the routing strategy
becomes more congestion-aware, the traffic is diverted from shortest path to
free (and longer) paths causing that both the mean speed of the vehicles and
the mean flux on the streets are largely increased with respect to the case
$\alpha=0$, since more vehicles are now able to reach their destinations
without being blocked in congested nodes. More interestingly, we observe in
Fig.~\ref{fig:fundamental}(b) that for $\alpha=2$ and $3$ the average
velocity $v$ does not decrease monotonically as a function of $L$ but it
reaches a maximum $v_{max}(\alpha)$ at some load $L^*(\alpha)$. However,
having a larger average velocity does not imply that the network is working in
a more efficient way. In fact, for large values of $\alpha$, the vehicles can
move faster running across longer paths at the expense of delaying the arrival
to their destinations. A measure of the efficiency of the routing
is the average number of routes $r$ completed by a vehicle during one hour.

The value of $r$ is reported in Fig.~\ref{fig:fundamental}(c) as a
function of $L$. The results indicate that, because of congestion, the
number of completed routes decreases when the number of vehicles in
the city increases. The precise dependence of $r$ on the load, is
related to the value of $\alpha$. For $\alpha = 0$, the average
vehicle speed has a very sharp drop as the load increases. For the
congestion-aware strategy with $\alpha=1$ the decrease is smoother
than for $\alpha = 0$, while for $\alpha=2$ ($\alpha=3$) the value of
$r$ is smaller than that of $\alpha=1$ when $L < 0.1$ ($L <0.14$), but
larger when $L > 0.1$ ($L > 0.14$). In practice, for a given load $L$,
the function $r(\alpha)$ shows a maximum at some $\alpha^*$ [see inset
in Fig.~\ref{fig:fundamental}(c)] . The value of $\alpha^*$ is seen to
increase with the load $L$, pointing out that the more congested the
network is, the more congestion-aware has to be the routing to reach
the optimal functioning.
On the other hand, the maximum value of $r$, $r(\alpha^*)$,
decreases with $L$.
Similar results as those shown for Bologna have been found for the
other cities studied. The best routing exponents $\alpha^*$
obtained for two realistic values of the vehicle density, namely
$L=0.06$ and $L=0.1$ \cite{chowdhury00}, are reported in
Table~\ref{table}. It is clear that the optimal value $\alpha^*$
depends strongly on the particular topology of the urban graph.

Even though a precise information on global congestion is in practice rarely
available, we finally study the case in which each vehicle knows exactly the
congestion in every link of the network. Namely, we compare the results
obtained with Eq. (\ref{eq:penalty}) with those obtained with the penalty
function
\begin{equation}
P_n = (d_{in} + d_{nt}) (1+{\left\langle c_{int}\right\rangle})^\alpha\;,
\end{equation}
where $\langle c_{int} \rangle$ accounts of the average road congestion along
the path from $i$ to $t$ passing by $n$. Therefore, we now project the effect
of the average congestion of the path (obtained from the global knowledge) on
the path distance.

The average number $r$ of completed routes per hour is reported in
Fig.~\ref{fig:fundamental}(d). The figure shows that the routing
with global knowledge does not perform much better than that of
Eq. (\ref{eq:penalty}). Additionally, at large loads, local
knowledge outperforms global knowledge in terms of $r$. Moreover,
when global information is taken into account, the optimal routing
$\alpha^*$ increases. This is related to the fact that, since
congestion in links close to vehicle location is always up-to-date
and therefore accurate, routing based on local congestion needs
lower values of $\alpha^*$ to divert vehicles on free streets.


{\em Conclusions.-} Congestion in transportation and communication
networks is a serious problem for both public goods and users
time. In this paper we have integrated the three essential
ingredients of vehicular traffic in urban settings. Namely the
graph structure of urban patterns, a cellular automata model for
vehicular dynamics along the links, and the use of a
congestion-aware routing inspired to data traffic in the Internet.
We thus have provided with a simple and feasible model where only
local information about traffic congestion is used to route
vehicles. Our results show how each individual agent can better
organize its motion based on its limited local knowledge of
traffic and, at the same time, achieve optimal performances at the
global system level. We have implemented this model in several
real cities with different structural properties showing how the
optimal vehicle routing strategy depends on the network topology.
Finally, we have shown that, counterintuitively, a (unfeasible)
routing based on a global knowledge of congestion is not suited
when the load of vehicles is large. The proposed routing model is
general enough to be applied to several types of human transport
networks. Moreover, the dynamic and distributed nature of the
model allows several applications to be built which may implement
the congestion-aware strategy to improve the available vehicle's
navigation systems.

{\em Acknowledgments.-} We are indebted to C. Mascolo and I. Leontiadis for
their valuable comments. V. Latora was supported by the Engineering and
Physical Sciences Research Council [grant number EP/F013442/1].


\begin{thebibliography}{99}
\bibitem{timestep} A time step is assumed to be be equal to 1 sec,
the minimum time of reaction of a driver.
\end{thebibliography}
\end{document}